\documentstyle[aps, draft, epsf, floats]{revtex}

\def\sm{Standard Model}

\def\up#1{^{(#1)}}
\def\alpbf{{\boldmath\alpha}}
\def\betbf{{\bf\beta}}
\def\mubf{{\bf\mu}}
\def\ocal{{\cal O}}

\begin{document}
\title{CP violation in gauge theories \\ \vskip - .25 in
\rightline{\parbox{1.1in}{\small UCRHEP-T284 \\ UM-TH-00-17}}}
\author{ Martin B. Einhorn }
\address{Randall Laboratory of Physics \\
University of Michigan \\
Ann Arbor, Michigan, 48019-1120\\
}
\author{ Jos\'{e} Wudka }
\address{
Department of Physics, University of California-Riverside \\
Riverside, California, 92521-0413
\\{\rm and}\\
Departamento de F\'\i sica Aplicada, CINVESTAV-IPN unidad M\'erida\\
M\'erida, Yucat\'an 97310, M\'exico
\vspace{.3cm}
}

\date{\today}

\maketitle

\begin{abstract}

We define the CP transformation properties of scalars, fermions and
vectors in a gauge theory and show that only three types of interactions
can lead to CP violation: scalar interactions, fermion-scalar
interactions and $ F \tilde F $ which is associated with the strong CP problem
and which involves only the gauge fields. For technicolor theories this
implies the absence of CP violation within perturbation theory.

\end{abstract}

\bigskip\bigskip

\paragraph{Introduction}
The origin of CP violation is one of the current important questions in
high-energy physics. There are three types of 
renormalizable interactions that are
commonly considered and that violate CP: Yukawa couplings, scalar
interactions and $ W \tilde W $-type terms associated with global chiral
anomalies. 
These are, in fact, the {\em only} CP-violating 
renormalizable interactions~\cite{rg}.
In this short note we consider some consequences of this fact
relevant for the observablity of {\em non}--Standard Model CP-violating 
effects; in addition we provide a simplified proof of the absence
of (perturbative)  CP-violating effects within the gauge sector of a 
general renormalizable theory with vectors and scalars.

Of course, the most popular extensions of the SM, viz., supersymmetric 
models, have scalar fields with Yukawa interactions and self-interactions 
so this result does not restrict them much.  However, in
view of this theorem, the {\it only} explicit CP-violating mechanism
that occurs in field theories {\it without} scalar fields is due to
global chiral anomalies.  Any other mechanism will be due to spontaneous
symmetry breaking associated with the formation of scalar condensates,
as are presumed to form in technicolor models.  Such theories
necessarily involve strong gauge interactions and so are difficult to
deal with, at least in a calculable manner.  However, if a sufficient
number of such scalars are present,  they can in principle lead to CP-
violating effects.\cite{cptc}   This contrasts with spontaneous parity
violation, which cannot occur in theories with only vector-like gauge-
interactions of fermions.\cite{vw}  Thus, it is the spontaneous breaking
of charge-conjugation that underlies this kind of CP-violation.  

As an application of this result we investigate the observability of
virtual CP-violating effects generated by physics beyond the standard
model in the case where the heavy physics is described by a weakly-coupled, 
natural gauge theory. In the following the label ``natural''
will be used in the technical sense: the smallness of coefficients are
protected by symmetries~\cite{thooft}. We argue that CP-violating
effects are very hard to distinguish from those produced by the Standard
Model so that, at least for this class of heavy theories, CP-violating
processes are {\em not} a sensitive probe of physics {\em beyond} the
Standard Model.~\footnote{These arguments presuppose that there is no
direct observation of the heavy excitations, that is, that the scale of
new physics is not directly accessible. In cases where this is not true
-- as, for example, occurs in some supersymmetric extensions of the
Standard Model -- the observability of CP-violating effects must be
determined on a case-by-case basis.}

For clarity we will first present the proof that there is a natural
definition of the CP transformation that reduces to the usual one and
for which all renormalizable gauge interactions are invariant. The second
part of the paper contains the phenomenological application of the result.

\paragraph{Properties of the group generators}
Although the definition of the CP transformations is not
unique~\cite{w.and.l}, the one provided below has the virtue of
generalizing the one used in the Standard Model. The simplest
expressions for the transformation of the fields are obtained when the
group generators are expressed in a Cartan basis. We denote by $H_i$ the
Cartan generators and by $ E_\alpbf $ the generator corresponding to the
root $ \alpbf $, then 
\begin{equation}
\begin{array}{ll}
\left[ H_i , H_j \right] = 0 & \qquad
\left[ E_\alpbf , E_{-\alpbf} \right] = \alpha_i H_i \\
\left[ H_i , E_\alpbf \right] = \alpha_i E_\alpbf & \qquad
\left[ E_\alpbf , E_\betbf \right] = N_{\alpbf , \betbf} E_{ \alpbf +
\betbf } \quad (\alpbf + \betbf \ \mbox{root}) ;
\end{array}
\end{equation}
we will not need the explicit expression for $ N_{\alpbf , \betbf} $,
only the property~\cite{gilmore}
\begin{equation}
N_{\alpbf, - \betbf} = - N_{- \alpbf , \betbf} .
\label{n.prop}
\end{equation}
We will denote the simple roots by $ {\bf a}_r $ and the fundamental
weights by $ \mubf_r $, then $ \mubf_r \cdot {\bf a }_s =
\delta_{rs} $.

We will show that for any irreducible representation(irrep) it is
possible to choose a basis where all the generators are real and such
that $E_{-\alpbf}^T = E_\alpbf $. To this end we note first that the
Cartan generators can taken to be Hermitian and diagonal, hence they are
real. It then follows that the complex conjugate $ E_\alpbf^* $
corresponds to the same root as $ E_\alpbf $ and therefore~\cite{georgi}
$ E_\alpbf^* = \lambda_\alpbf E_\alpbf $ for some complex numbers $
\lambda_\alpbf $ obeying
\begin{equation}
\left| \lambda_\alpbf \right| = 1 , \qquad \lambda_\alpbf \lambda_\betbf
= \lambda_{ \alpbf + \betbf } \ \mbox{when}\ \alpbf + \betbf \ \mbox{is
a root}
\label{lam.prop}
\end{equation}
(the second result follows from the commutators); as a special case we
have $ \lambda_\alpbf \lambda_{ - \alpbf } = 1 $. Noting now that $
E_\alpbf^\dagger = E_{ - \alpbf } $ we obtain
\begin{equation} 
E_{- \alpbf} = \lambda_\alpbf E_\alpbf^T
\end{equation}
where $T$ denotes transposition.

Using the expansion of $\alpbf $ in terms of the simple roots $ \alpbf =
\sum_r c_r {\bf a}_r $ together with  (\ref{lam.prop}) implies 
\begin{equation}
\lambda_\alpbf = \prod_r \lambda_{ {\bf a}_r }^{c_r }
\end{equation}
Furthermore, since $ |\lambda| = 1 $ we can write $\lambda_{ { \bf a }_r
} = \exp \left( i \phi_r \right) $ which we use to define the unitary
matrix 
\begin{equation}
U = \exp \left( { i \over 2} \sum_{r, i}\phi_r \; H_i \cdot \mubf_r^i
\right) 
\end{equation}
where, as before, $ \mubf_r $ denote the fundamental weights.

We now use $U$ to generate a new basis. We clearly have $ H_i' = U H_i
U^\dagger = H_i $ while the roots are rescaled appropriately,
\begin{equation}
E_\alpbf'  = U
E_\alpbf U^\dagger = \exp \left( + { i \over 2 } \sum_r \phi_r c_r
\right) E_\alpbf = \prod_r \lambda_{ {\bf a}_r }^{ c_r /2 } E_\alpbf =
\lambda_\alpbf^{1/2} E_\alpbf. 
\end{equation}
Then
\begin{equation}
E_\alpbf'{}^* =
\left( \lambda_\alpbf^{1/2} \right)^* \lambda_\alpbf E_\alpbf = 
\sqrt{ \lambda_\alpbf } E_\alpbf = E_\alpbf'
\end{equation}
In this basis all the generators are real and, dropping the primes,
\begin{equation}
H_i = H_i^* = H_i^T \qquad
E_\alpbf = E_\alpbf^* = E_{- \alpbf}^T 
\label{e.transp}
\end{equation}
proving our assertion~\cite{rg}.

\paragraph{CP transformation for the fields}
We now consider a gauge theory with gauge fields $W_\mu$, fermions  $
\psi $  assumed to be contained in a single multiplet that in general
transforms  reducibly, and scalars $ \phi $ also lumped into one
multiplet. We define the CP transformation of the fields as follows,
\begin{equation}
\begin{array}{ll}
\psi \to  C \psi^*  & \qquad W_\mu\up i     \to  - W\up i {}^\mu  \\
\phi \to    \phi^*  & \qquad W_\mu\up\alpbf \to  - W\up{ - \alpbf }
{}^\mu 
\label{field.transf}
\end{array}
\end{equation}
where $C$ denotes the charge conjugation matrix (acting on the Lorenz
indices of $ \psi $). It is understood that the arguments of the fields
become their $P$ transforms and that the fields are in a basis whose
gauge generators obey (\ref{e.transp}). It is easy to see that for the
standard model (\ref{field.transf}) reduces to the usual expressions,
in particular the change in sign of the root $\alpbf$ corresponds to
a replacement of positive charged fields by negative charged ones.

With these definitions and using (\ref{e.transp})
the currents are seen to obey
\begin{eqnarray}
j_{ L/R \, \mu } \up i = \bar \psi H_i \gamma_\mu P_{ L/R } \psi 
& \stackrel{CP}{\longrightarrow} &- j_{ L/R } \up i {}^\mu  \nonumber \\
j_{ L/R \, \mu } \up\alpbf = \bar \psi E_\alpbf \gamma_\mu P_{ L/R }
\psi 
& \stackrel{CP}{\longrightarrow} & - j_{ L/R } \up{-\alpbf} {}^\mu
\nonumber \\
\end{eqnarray}
which implies that the fermionic kinetic Lagrangian $ \bar \psi \not
\!\! D \psi $ is invariant.

Denoting the curvature associated with $W_\mu$ by $ W_{ \mu \nu} $ and
using
(\ref{n.prop}) and (\ref{field.transf}), we get 
\begin{eqnarray}
W_{ \mu \nu } \up i \to - W \up i {}^{ \mu \nu }   \qquad
W_{ \mu \nu } \up\alpbf \to - W \up{ - \alpbf } {}^{ \mu \nu }   
\end{eqnarray}
so that the gauge Lagrangian $ -{1\over4} W_{ \mu \nu } ^2 $ is also
invariant~\footnote{The term $ W_{\mu \nu } \tilde W^{ \mu \nu} $ is
odd under CP but it is a total derivative and so has no effects to any
finite order in perturbation theory.}.

An explicit mass term for the fermions (if present) transforms 
according to 
\begin{equation}
\bar \psi M \psi \stackrel{CP}{\longrightarrow} \bar \psi M^T \psi. 
\end{equation}
where we have assumed that possible terms proportional to $ \gamma_5 $
have been eliminated through a chiral rotation; the Jacobian associated
with such a transformation adds a term proportional to $ W \tilde W $ to
the Lagrangian and has no perturbative effects~\cite{fuj}. Within each
irreducible representation the mass submatrix is proportional to the
identity. If there are two multiplets $ \psi_1 $ and $ \psi_2 $
transforming according to the same irreducible representation we can
replace them by arbitrary (unitary) linear combinations, and so we can
choose $M$ to be diagonal and real. It follows that this term is CP
invariant. In contrast, terms of the form $ \bar \psi \psi \Gamma \phi $
(where $ \Gamma $ denotes the Yukawa coupling matrix) are not, in
general, CP invariant.

Finally, using the fact that the $H_i $ are real and diagonal together
with (\ref{e.transp}) we obtain
\begin{equation}
D_\mu \phi \to \left( D^\mu \phi \right)^*
\end{equation}
which shows that the scalar kinetic Lagrangian $ (D_\mu \phi )^\dagger
\; D^\mu \phi $ is also an invariant. Just as for the fermion mass we
can choose the scalar mass  matrix $ {\cal M }^2 $ to be real and
diagonal. It then follows that the term $ \phi^\dagger {\cal M}^2 \phi $
will be invariant under CP.

These results show that, in the absence of spontaneous symmetry
breaking, the only renormalizable terms that can violate CP in the Lagrangian are the
fermion-scalar interactions (Yukawa couplings) and the scalar potential
$V$.

\paragraph{Virtual CP-violating effects}

We now imagine that the \sm\ is the low-energy limit of a weakly-coupled
natural gauge theory. In this case the corrections produced by virtual
heavy physics effects to low-energy processes are determined by all
gauge-invariant local operators involving the Standard Model fields.
Leading effects are produced by operators with the largest coefficients.
For the class of theories under consideration these are operators
generated at tree-level \cite{aew}. 

We apply the previous result to this case as follows: we first isolate the
effective operators that violate CP and whose coefficients are not suppressed
by loop effects or constrained by existing data. These operators are of
interest because, if present, they would provide the strongest non-standard
CP-violating signals. We then describe the types of heavy physics that can
generate these operators. At this point we can restrict ourselves to the case
where the {\em heavy} physics is described by a renormalizable theory.
Indeed, if the
heavy scale is denoted by $\Lambda $, any operator of dimension higher than 4
present at this scale would be generated by interactions whose scale
$\Lambda'$
is  larger than $\Lambda $, and the corresponding coefficient will be 
suppressed by an 
inverse power of $\Lambda'$; at low energies the corresponding effects
would be suppressed by a power of $ \Lambda / \Lambda' \ll 1$ and can be
ignored.

The specific form of these operators depends, of course, on the spectrum
of light particles, we will assume that these include the usual \sm\
fermions and gauge bosons and, in addition, one scalar doublet. For this
case the list of all tree-level-generated operators is known \cite{aew}
and, using the above transformation rules, those which are not CP
invariant can be isolated; they are~\footnote{We denote the left-handed
quark and lepton doublets by $q$ and $ \ell$; right-handed up and down
quarks are denoted by $u$ and $d$, right-handed charged leptons by $ e$;
the scalar doublet is labeled $ \phi$. The generators of $SU(3)$ are
denoted by $ \lambda^A$, the ones for $SU(2)_L$ are the usual Pauli
matrices $\tau^I,$ and $ \varepsilon = i \tau^2 $. $D_\mu$ denotes the
covariant derivative and $v$ the scalar doublet vacuum expectation
value.}
\begin{eqnarray}
I:&& \left\{ \begin{array}{lll}
\left(\bar\ell e\right)\left(\bar d q\right)-\hbox{h.c.} &
\left(\bar q u\right) \varepsilon\left(\bar q d\right)-\hbox{h.c.} &
\left(\bar q \lambda^A u\right) \varepsilon\left(\bar q
\lambda^Ad\right)-\hbox{h.c.}\\
\left(\bar \ell e\right) \varepsilon\left(\bar q u\right)-\hbox{h.c.} &
\left(\bar \ell u\right) \varepsilon\left(\bar q e\right)-\hbox{h.c.}
& \\
\end{array} \right.\cr
II:&& \left\{ \begin{array}{lll}
\left(|\phi|^2 -v^2 \right) \left(\bar\ell e \phi-\hbox{h.c.}\right) &
\left(|\phi|^2 -v^2 \right) \left(\bar q u \tilde\phi-
\hbox{h.c.}\right) &
\left(|\phi|^2 -v^2 \right) \left(\bar q d\phi-\hbox{h.c.}\right) \\
\end{array} \right.\cr
III:&& \left\{ 
\left(\bar f \gamma^\mu f \right) \partial_\mu \left(|\phi|^2 -v^2
\right)  , \quad f=u,d,e,q,\ell  \right.\cr
IV:&& \left\{ \begin{array}{l}
\ocal_1=\left(\phi^\dagger\tau^I\phi\right) D^{IJ}_\mu \left(\bar q
\gamma^\mu \tau^J q\right) \\
\ocal_2=\left(\phi^\dagger\tau^I\phi\right) D^{IJ}_\mu \left(\bar \ell
\gamma^\mu \tau^J \ell\right) \\
\ocal_3=\left(\phi^\dagger \varepsilon D_\mu\phi\right) \left(\bar u
\gamma^\mu d \right) -\hbox{h.c} \\
\end{array} \right.
\label{ops}
\end{eqnarray}
each of the fields has, in principle a family index which we have
suppressed.

Operators not included in the above list may arise from loop
corrections, and so will have coefficients suppressed by a factors of $
g^2/(4\pi)^2 $, where $g$ is some generic coupling constant. Their
corresponding CP-violating effects are typically down by a factor $ 
\sim g^2 v^2/(4 \pi \Lambda)^2 $ where $v\sim 246$~GeV denotes the vacuum
expectation value of the Standard Model scalar doublet (we assume weakly
interacting heavy physics). On the other hand, Standard Model CP
violating effects are suppressed by a small factor $ \sim 4 \times 10^{-
5} $ proportional to a product of CKM matrix elements~\cite{cp.sm}, so
that the Standard Model effects will dominate for $ \Lambda > 3$~TeV.
Moreover it is possible for the underlying theory to have additional
suppression factors not mandated by the gauge symmetry. So, even though
loop generated operators may produce a wider variety of CP-violating
signatures than those in (\ref{ops}), we expect the corresponding
effects to be, at best, marginally inside the current sensitivity. For
moderately heavy physics ($ \Lambda > 3$~TeV) these small effects need
to be disentangled from similar Standard Model contributions. We
therefore conclude that the highest sensitivity to heavy CP-violating
physics is obtained in reactions involving the operators in (\ref{ops})
to which we now turn.

The 4-fermion operators of group $I$ in (\ref{ops}) contribute (at one
loop) to the strong CP parameter $ \theta $ and so their coefficients
are strongly constrained (assuming no cancelations). Additional
constraints can be obtained for fermions in the first generation from
meson decays. Moreover, several of these operators (depending on the
generation to which each of the fields belongs) will contribute
radiatively to fermion masses, so that their coefficients will be
suppressed by the (smallest) corresponding Yukawa coupling for natural
heavy physics.

Operators involving scalars and fermions that violate chirality (group
$II$ in (\ref{ops})) also contribute to the fermion masses and their
coefficients are correspondingly suppressed in natural theories (except
possibility for the ones contributing to the top-quark mass). Note also
that operators in groups $II$ and $III$ in  \ref{ops} can probed only in
processes involving real Higgs particles.

Thus the only unsuppressed operators that can be probed in current and
near-future experiments and that do not involve scalars are those in
group $IV$. These take the form (in unitary gauge)
\begin{eqnarray}
\ocal_1 &\to& -{i g v^2\over \sqrt{2}} \left(\bar u_L \not\!\!W^+ d_L -
\hbox{h.c.}\right) + \hbox{terms~with~scalars}\cr
\ocal_2 &\to& -{i g v^2\over \sqrt{2}} \left(\bar \nu_L \not\!\!W^+
e_L -\hbox{h.c.}\right) + \hbox{terms~with~scalars}\cr
\ocal_3 &\to& -{i g v^2\over \sqrt{8}} \left(\bar u_R \not\!\!W^+ d_R -
\hbox{h.c.}\right) + \hbox{terms~with~scalars}
\label{unit}
\end{eqnarray}

The Higgs-independent terms in $\ocal_{1,2}$ are indistinguishable from
terms already present in the \sm. The Higgs-independent terms in the
operators $ \ocal_1$ generate a shift in the CKM matrix elements $
V_{ij} \to V_{ij} + g(v/\Lambda)^2 v_{ij} + O(1/\Lambda^4) $ which is
not necessarily unitary. 

For massless neutrinos a basis of lepton fields can be chosen so that
the operators $ \ocal_2 $ do not mix families. The effects of the Higgs
independent terms then correspond to a violation of universality in the
couplings of the leptons to the $W$ boson. Current data~\cite{pdg}
require $ \Lambda > 2.2$~TeV (at 1$\sigma$ assuming the coefficient of
the operator is $1/\Lambda^2$). For massive neutrinos the effects of
these interactions correspond to a shift in the leptonic CKM matrix.
Finally $\ocal_3$  corresponds to a right-handed quark current and
contributes to the $W$ mass and $\tau$ decay; existing data implies $
\Lambda > 500 $~GeV~\cite{pdg}.

\nobreak
\begin{figure}
\vbox to 3 in{\epsfxsize=5 in\epsfbox[-100 -500 512 292]{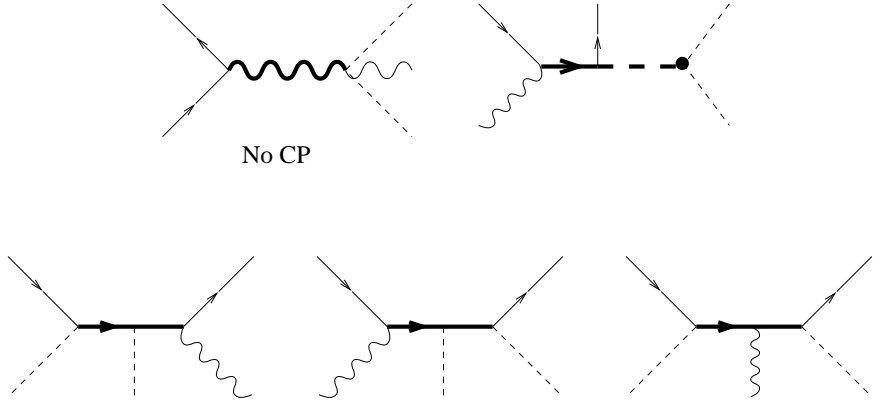}}
\caption{Tree-level graphs that generate
effective CP violation operators $ \ocal_{1,2,3}$.
Solid straight lines denote fermions, dashed lines denote scalars, wavy
lines denote vector bosons. All thick lines represent heavy particles,
the heavy dot denotes a super-renormalizable (SR) vertex $ \sim
\Lambda $.} 
\label{f6}
\end{figure}

The possible tree-level graphs of the underlying theory that can
generate $ \ocal_{1,2,3}$ were determined in \cite{aew} and are shown in
Fig. \ref{f6}. In the following we call $ \Phi $ a generic heavy scalar,
$F$ a heavy fermion, $f$ a light (Standard Model) fermion and $ \phi $
the Standard Model scalar doublet. Note that in view of the
result proved above and in \cite{rg} we can immediately dismiss those
graphs containing only gauge interactions as indicated in the figure.
Moreover, knowing the structure of the graphs allows us to draw
a number of general properties that the heavy fermions and
scalars must satisfy in order to contribute to the operators of
class $IV$.

Consider first the diagram containing a heavy scalar $\Phi$, which must
be an $SU(2)$ triplet and a $U(1)_Y$ singlet; one can also show that in
order to maintain $v\ll \Lambda $ while having a $\Phi$ mass of order $
\Lambda $, the triplet $\Phi$ must get a vacuum expectation value of
(at most) order $v^2/\Lambda $. Given the previous transformation laws the
$\phi\phi\Phi$ coupling is CP conserving, hence the CP-violating
properties of this graph can come only form the $F f\phi$ couplings. For
this coupling to exist the heavy fermions must be $SU(2)$ doublets, so
that this graph cannot contribute to $ \ocal_3$.

Similar considerations apply to the remaining graphs that involve only
heavy fermions. All involve a heavy fermion doublet (under $ SU(2) $)
and only the last graph can generate contributions to $ \ocal_3 $. It
follows that for natural, weakly-coupled heavy physics, observable CP
violating interactions are generated by heavy fermions through their
mixing with Standard Model fermions and/or through their Yukawa
couplings with the Standard Model scalar doublet. In addition the Yukawa
couplings of the form $F f \phi$ are suppressed by a factor of the $f$
mass (having assumed naturality) and so the most significant effects can
be expected in processes involving the third generation of Standard
Model fermions.

We conclude that for natural, weakly-coupled heavy physics the most
significant CP-violating effects appear through violation of the
unitarity relations in the CKM matrix for quarks (and, for massive
neutrinos, for leptons also), in violations of universality in the
coupling of the $W$ to the fermions, or in the appearance of a right-
handed charged current; these effects can be most noticeable when
involving the top quark.  We also note that stronger CP effects may be
observed provided in reactions involving scalars. The above arguments
also imply that existing data on right-handed currents can provide
indirect bounds on the scale of heavy, CP-violating physics.

\end{document}